\documentclass[twocolumn,amsmath,amssymb,aps,pre,showpacs,floatfix]{revtex4}
\usepackage{graphics}

\begin{document}

\title{ Fluctuation formula for nonreversible dynamics in
        the thermostated Lorentz gas}

\author{M. Dolowschi\'ak} \email{dolowsch@szerenke.elte.hu}
\author{ Z. Kov\'acs} \email{kz@garfield.elte.hu}

\affiliation{Institute for Theoretical Physics, E\"otv\"os University, 
          Pf.\ 32, H--1518 Budapest, Hungary}

\begin{abstract}
We investigate numerically the validity of the Gallavotti-Cohen 
fluctuation formula in the two and three dimensional periodic 
Lorentz gas subjected to constant electric and magnetic fields 
and thermostated by the Gaussian isokinetic thermostat. 
The magnetic field breaks the time reversal symmetry, and by 
choosing its orientation with respect to the lattice one can have 
either a generalized reversing symmetry or no reversibility at all.
Our results indicate that the scaling property described by the 
fluctuation formula may be approximately valid for 
large fluctuations even in the absence of reversibility.
\end{abstract}


\maketitle

The Lorentz gas (LG) thermostated by a Gaussian isokinetic (GIK) thermostat
is one of the most popular models in the study of the relationship between
transport properties and chaotic behaviour in nonlinear dynamical systems.
Since the microscopic dynamics of the LG is chaotic, and
on a sufficiently long time scales it possesses a well defined 
macroscopic transport coefficient, it can be used to study the connection 
of microscopic chaos and macroscopic nonequilibrium behaviour.

The so-called fluctuation formula (FF) has first been 
observed numerically in a system of thermostated fluid particles 
undergoing shear flow \cite{shearflow}. 
In that model, trajectory segments violating the second law of 
thermodynamics were found with probabilities 
exponentially smaller than those of trajectory 
segments associated with normal thermodynamical behaviour. 
More precisely, 
let $\xi_\tau(t)$ denote the {\em entropy production rate} $\xi$ 
averaged over a time interval of length $\tau$ centered around time $t$:
$\xi_\tau(t)= \frac{1}{\tau}
      \int\limits_{-{\tau}/{2}}^{{\tau}/{2}}\xi(t+t')\,\mathrm{d}t'$, 
and let us consider it as a probabilistic variable. 
Then its statistical properties in a steady state can be 
characterized by a probability density $\Xi_\tau(x)$. 
The fluctuation formula states \cite{orig} that this probability 
density has the following property:
\begin{equation}
  \label{flform}
  \lim_{\tau\to\infty}
        \frac{1}{\tau} \ln \frac{\Xi_\tau(x)}{\Xi_\tau(-x)}=x\,. 
\end{equation}
One of the interesting features of the FF is that it seems to be valid 
in systems far from equilibrium, not just for vanishing external fields. 

After discovering the formula numerically, 
analytical results were obtained about its validity  
in deterministic systems like
transitive Anosov systems \cite{dynens} and
special reversible maps \cite{baker,dorfmannbook}.
In the proofs of these theorems, the {\em time reversibility} of the 
system plays a key role \cite{stocha}.
Nevertheless, proving fluctuation theorems under more general
conditions seems to be exceedingly difficult.
In this context, even relatively simple systems like the LG, 
with or without magnetic field, 
seem to be out of reach for the existing analytical techniques.
It is also unclear
how the FF should look like in systems with nonreversible dynamics 
\cite{rem-windtree}.
Consequently, reliable numerical results for such models may provide
valuable hints in the search for more sophisticated theoretical approaches.

The field driven Lorentz gas consists of a charged particle subjected to 
an electric field moving in the lattice of elastic scatterers.
For the sake of simplicity, we take a square or cubic lattice in our study,
depending on the dimensionality of the system.
Due to the applied electric field, one must use a thermostating mechanism 
to achieve a steady state in the system.
Such a tool is the Gaussian isokinetic thermostat which preserves 
the kinetic energy of the particle; 
for a review see e.g. \cite{thermostats} and further references therein.
We will also apply a constant external magentic field to control
the reversibility of the dynamics.

Throughout our work we use dimensionless variables.
We choose the units of mass and electric charge to be equal to 
the mass and electric charge of the particle, so we have $m=q=1$ 
in our model.
The unit of distance is taken to be equal to the radius of scatterers ($R=1$),
and the unit of time is chosen to normalize the magnitude of particle
velocity to unity.
Let ${\bf q}=\left(q_1,\dots,q_n\right)$ denote the position and
${\bf p}=\left(p_1,\dots,p_n\right)$ the momentum of the particle
in the $n$-dimensional space ($n=2$ or 3).
Due to the normalization, $\left|{\bf p}\right|=1$.
The phase space variable of the system is
${\mathbf \Gamma}=\left({\bf q},{\bf p}\right)$;
it is transformed abruptly at every 
elastic collision and evolved smoothly by the differential equation
\begin{eqnarray}
  \label{diffeq}
   \dot{\bf q} & = & {\bf p}       \nonumber \\
   \dot{\bf p} & = & {\bf E} + {\bf p}\times{\bf B} - \alpha {\bf p}
\end{eqnarray}
between them. 
Here $\alpha$  is called the {\em thermostat variable}, 
while ${\bf E}$ and ${\bf B}$ are constant vectors
playing the roles of the external electric and magnetic fields, 
respectively.
The GIK thermostat corresponds to the choice $\alpha={\bf E}{\bf p}$
in Eq.~(\ref{diffeq}).
For $n=2$, ${\bf B}$ is thought to be perpendicular 
to the plane of motion given by the directions of 
${\bf E}$ and ${\bf p}$.
We note that Eq.~(\ref{diffeq}) is dissipative, but
for ${\bf B} = 0$ it has also time reversal symmetry.

Dissipation can be measured by the phase space contraction rate 
$\sigma$.
It can be computed by taking the divergence of the right-hand side  of 
Eq.~(\ref{diffeq}):
\begin{eqnarray}
  \label{sigma}
   \sigma= - \mbox{div}\, {\bf \dot\Gamma} =-\left(n-1\right)\alpha\,,
\end{eqnarray}
and it can be shown (see e.g. \cite{epr}) that in our case
\begin{eqnarray}
\label{sigmaxi}
       \sigma(t)&=&\xi(t) .
\end{eqnarray}
We note that the validity of this identity does depend on the choosen
model and cannot be treated as a general property  
\cite{thermostats,nosehoover,detscatt}.

The notion of reversibility \cite{revers}, 
an extension of time reversal symmetry, 
can be formulated in terms of the phase space flow
${\bf \Phi}^{t}$ defined by ${\bf \Gamma}(t)={\bf \Phi}^{t}{\bf \Gamma}_{0}$.
We say that the flow is reversible,
if there exists a map ${\bf G}$ which is an involution 
(i.e ${\bf G}^2$ is the identity) and bracketing the flow by ${\bf G}$
reverses the direction of time:
\begin{eqnarray}
\label{rev}
   {\bf G}{\bf \Phi}^{t}{\bf G}={\bf \Phi}^{-t}.
\end{eqnarray} 
Time reversal symmetry is a special case of reversibility 
with a particular choice of the involution:  
${\bf G}_{0}({\bf q},{\bf p})=({\bf q},{\bf -p})$.

In the LG, reversibility depends on the directions of the field
vectors relative to each other and the lattice.
It can be checked easily that our system is time reversible if
${\bf B}={\bf 0}$, and it is not otherwise. 
In Ref.~\cite{cpr}, we have also shown that the system is still 
{\em reversible} for ${\bf B} \neq {\bf 0}$ if the plane containing 
${\bf E}$ and ${\bf B}$ is a symmetry plane of the lattice.
Since the transformation ${\bf G}={\bf M}{\bf G}_0$ (where 
${\bf M}$ is a mirroring of ${\bf q}$ and ${\bf p}$ with respect 
to the plane containing ${\bf E}$ and ${\bf B}$) satisfies 
Eq.~(\ref{rev}), 
the smooth flow is always reversible. 
This means that the reversibility of the full dynamics including
the collisions requires 
that the invariant plane of ${\bf M}$ be a symmetry plane of the 
lattice \cite{cpr}.
In the two dimensional case this is simplified to the condition
that ${\bf E}$ has to be contained by the symmetry plane of the lattice. 

The goal of our numerical simulations was to measure $\Xi_\tau(x)$ with
a precision which is sufficient to check the validity of the fluctuation
formula. 
Due to Eq.~(\ref{sigmaxi}), $\Xi_\tau(x)$ could be measured by 
periodically computing 
$\sigma_\tau$ along a particle trajectory and making a histogram of 
these data. 
The disadvantage of this method is that the range of possible
$\sigma_\tau$ values depends on the strength of the electric field.
Instead we may introduce the quantity
\begin{eqnarray}
  \label{pitau}
  \pi_\tau(t)= \frac{1}{\tau}
 \int\limits_{-\frac{\tau}{2}}^{\frac{\tau}{2}} 
               {\bf n_E}{\bf p}(t+t^{'})\,\mathrm{d}t^{'},   
\end{eqnarray}
where ${\bf n_E}$ denotes the unit vector paralell to ${\bf E}$. 
Since the magnitude of ${\bf p}$ is unity, $\pi_\tau$ always satisfies 
$\pi_\tau\in[-1,1]$.
By making a histogram of the periodically measured values of $\pi_\tau$, 
one gets an approximation of its probability density $\Pi_\tau(x)$. 
Since the two probabilistic
variables satisfy $\sigma_\tau=(n-1)\,E\pi_\tau$, the connection of the 
probability densities is
\begin{eqnarray}
  \label{XiPi}
   \Xi_\tau(x)=\frac{1}{(n-1)E}\,\Pi_\tau\left(\frac{x}{(n-1)E}\right).
\end{eqnarray}
Then we can rewrite the fluctuation formula as
\begin{eqnarray}
  \label{flformPi}
  \lim_{\tau\to\infty}
  \frac{1}{(n-1)E}\frac{1}{\tau}\,\ln \frac{\Pi_\tau(x)}{\Pi_\tau(-x)}=x\,. 
\end{eqnarray}

\begin{figure}
\scalebox{0.3}{\rotatebox{-90}{\includegraphics{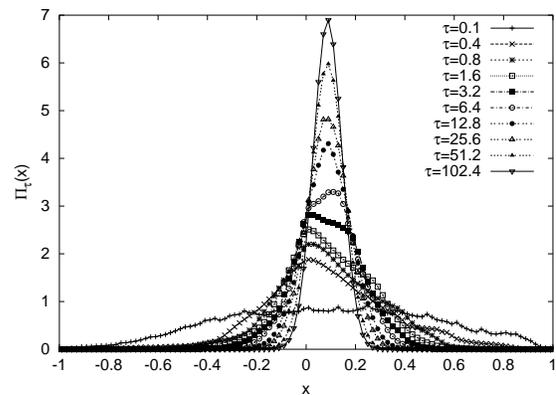}}}
\caption{The probability density $\Pi_\tau(x)$ for a 2D configuration,
where ${\bf E}=(0.5,0.8)$ and $|{\bf B}|=0.2$.
The distance between the centers of the scatterers is $d=2.1$;
the number of collisions is $1.6\times10^{8}$, while the average 
time between two collisions is $\approx 0.6$.
These data are similar throughout all examples presented in this paper.
}
\label{hist}
\end{figure}

\begin{figure}
\scalebox{0.3}{\rotatebox{-90}{\includegraphics{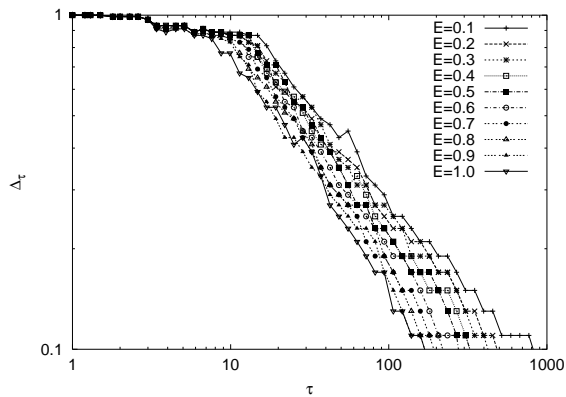}}}
\caption{The dependence of $\Delta_\tau$ on $\tau$ for different field
strengths in a 2D configuration with ${\bf B}=0$. 
The direction of ${\bf E}$ is parallel with $(5,8)$ but its magnitude varies.
The curves appear to be linear in the dominant region on the
log-log plot, suggesting a power law dependence on $\tau$.
This behavior seems to be valid for other configurations as well,
no matter they are reversible or not.
}
\label{delta}
\end{figure}

At a first glance, $\Pi_\tau(x)$ behaves similarly in all cases: 
as $\tau$ grows, $\Pi_\tau(x)$ becomes more and more 
concentrated around its mean value. 
This typical shape is shown in Fig.~\ref{hist}. 
It can be noticed that the curve looks like a Gaussian, 
although it is clear that it must be different due to the finite
range of $x$ \cite{rem-Gauss}.
In a separate paper \cite{next}, we will deal with the properties of 
this distribution in more details.
In order to visualize the fluctuation formula, we introduce the quantity
\begin{eqnarray}
\label{Dtau}
    D_\tau(x)=\frac{1}{(n-1)E}\frac{1}{\tau}\,
         \ln \frac{\Pi_\tau(x)}{\Pi_\tau(-x)}
\end{eqnarray}  
that must exactly be linear with a slope 1 in the $\tau \to \infty$ limit 
if the fluctuation formula is valid.
We will investigate for different configurations of the LG how well 
$D_\tau(x)$ approaches this behavior in numerical simulations.
Due to the fact that we have a finite number of data points 
coming from a numerical trajectory of finite length, 
our conclusions concerning $D_\tau(x)$ and thus the fluctuation formula
are, of course, limited to an interval 
$[-\Delta_\tau,\Delta_\tau]$ with $\Delta_\tau \le 1$.
In practice, if the extremal $\pi_\tau$ values in a series
of $N$ measured data were $\pi_{min}$ and $\pi_{max}$,
then we identified  $\Delta_\tau$ with $\min (-\pi_{min}, \pi_{max})$.
It is easy to check that the probability of observing $\pi_\tau$ values
outside this interval in another series is in the order of $1/N$.
Fig.~\ref{delta} shows the $\tau$ dependence of $\Delta_\tau$ for 
different field strengths with $N$ fixed.
For our simulations, $N$ was chosen to be $10^9$, which
means that if $D_\tau(x)$ is found to be linear on 
$[-\Delta_\tau,\Delta_\tau]$ with slope 1, 
then it can be interpreted as the fluctuation formula is valid
for fluctuations with probabilities larger than $10^{-9}$.

In the rest of the paper, we present our numerical results for the 
GIK thermostatted LG both in two and three dimensions, 
with various values of the external fields.
We focus on the question whether nonreversible dynamics leads to different 
scaling in the fluctuations than the one found in reversible systems.
As a general rule, we have not found any difference between time
reversal symmetric cases (i.e., with ${\bf B} = 0$) and reversible
ones.
Indeed, time reversal symmetry can be replaced in the known 
fluctuation theorems by general reversibility without affecting
their validity, since the proofs do not make use of the special
form of the involution ${\bf G}_0$.
We note that we have tested the different dynamical cases with several
choices for the field stregths and could not find significant deviations
in the observed behavior as long as we stayed within the ergodic region of the 
parameter space.
\begin{figure}
\scalebox{0.3}{\rotatebox{-90}{\includegraphics{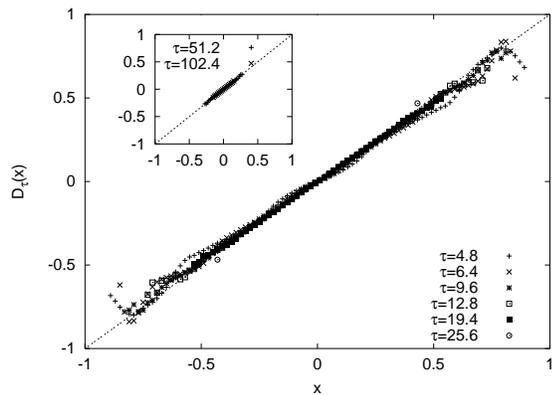}}}
\caption{$D_\tau(x)$ for a time reversible configuration (${\bf B}=0$) 
in 2D with ${\bf E}=(0.5,0.8)$. 
The inset shows that for higher $\tau$ values, $D_\tau(x) \approx x$ on 
$[-\Delta_\tau,\Delta_\tau]$. The inset has the same axes as the figure.
}
\label{2drev}
\end{figure}
\begin{figure}
\scalebox{0.3}{\rotatebox{-90}{\includegraphics{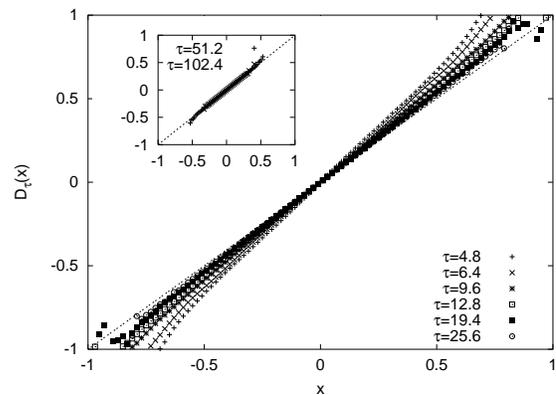}}}
\caption{$D_\tau(x)$ for a time reversible configuration in 3D,
with ${\bf E}=(0.05,0.1,0.15)$ and ${\bf B}={\bf 0}$. 
The inset shows that $D_\tau(x)$ converge to $x$ as $\tau$ gets larger.
The axes of the inset are the same as in the figure.
}
\label{3drev}
\end{figure}
In Figs.~\ref{2drev} and \ref{3drev}, we plot $D_\tau\left(x\right)$
for reversible dynamics in two and three dimensions (2D and 3D), 
respectively.
The fluctuation formula appears to be valid in both cases;
the convergence to the linear limit, however, seems to be different in them. 
For 2D, the $D_\tau\left(x\right)$ curve has deviations, 
decreasing in size with $\tau$ increasing, from the linear shape, 
while for 3D, $D_\tau\left(x\right)$ exhibits strongly linear
behavior with slopes approaching 1 as $\tau$ increases.
It is worth noting that the latter convergence can also be observed
in the 2D {\em random} LG  \cite{unpub}.

\begin{figure}
\scalebox{0.3}{\rotatebox{-90}{\includegraphics{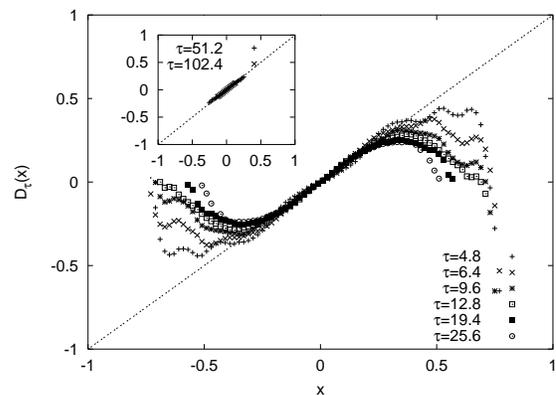}}}
\caption{$D_\tau(x)$ for a nonreversible configuration in 2D,
with $|{\bf B}|=0.2$ and ${\bf E}=(0.5,0.8)$.
It seems that for lower $\tau$ values $D_\tau(x)$ has a breakoff from
the linear curve around $x\approx\pm0.3$, but the inset shows that 
for higher $\tau$ values, $D_\tau(x)$ behaves quite similarly
to the reversible case of Fig.~\ref{2drev}.  
We note that this is the same configuration as the one used for 
Fig.~\ref{hist}.
The inset has the same axes as the figure.}
\label{2dnonrev}
\end{figure}

\begin{figure}
\scalebox{0.3}{\rotatebox{-90}{\includegraphics{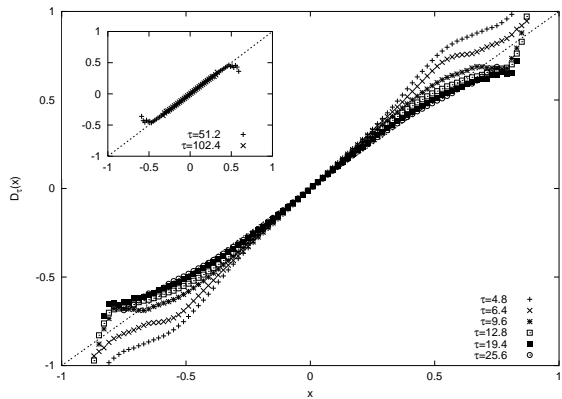}}}
\caption{$D_\tau(x)$ for a nonreversible configuration in 3D,
with ${\bf E}=(0.05,0.1,0.15)$ and ${\bf B}=(0.16,-0.06,0.04)$.
The axes of the inset are the same as in the figure.}
\label{3dnonrev}
\end{figure}

Our results for the nonreversible versions 
are shown in Figs.~\ref{2dnonrev} and \ref{3dnonrev}.
The most striking difference compared to the reversible cases is the
fact that there seems to be a {\em cubic} term present in 
$D_\tau\left(x\right)$ that does not disappear for larger $\tau$ values.
This term leads to a breakoff from the diagonal line for $|x| \geq 
x_c \approx 0.3$,
which means that there can be deviations from the fluctuation formula 
for {\em large} fluctuations. 
The slope of the linear part, however, is still 1 in the large $\tau$
limit, so the FF can be a good approximation for
small to moderate size fluctuations.
The fact that the region of validity of the FF does not shrink
considerably for larger $\tau$ values suggests that the
coefficient of the cubic term in $D_\tau (x)$ may have only weak 
dependence on $\tau$.
This also means that as the distribution $\Pi_\tau(x)$ is concentrating
around its mean value for increasing $\tau$ values, the total statistical
weight of the large fluctuations that are not covered by the linear
regime is decreasing.
In other words, the FF becomes more and more valid in a probabilistic
sense as $\tau \to \infty$, since the larger fluctuations become
less and less likely in that limit.

We may conclude that the FF appears to be valid  in the 
GIK thermostated LG with reversible dynamics, both in two and three 
dimensions.
For nonreversible dynamics, we have found indications that the
FF may still describe the scaling properties of fluctuations in a 
moderate size regime, although for large fluctuations there are
clear deviations from it due to higher order terms
in $D_\tau\left(x\right)$.
The fact that the slope of the linear part in the scaling behavior is
the same in reversible and nonreversible cases suggests a
kind of robustness for the FF in the thermostated LG.
It would be interesting to see if this remains valid in other
nonequlibrium systems with nonreversible dynamics.

\begin{acknowledgments}
The authors are grateful to Tam\'as T\'el for fruitful discussions
and a careful reading of the manuscript.
This work was supported by the Bolyai J\'anos Research Grant of
the Hungarian Academy of Sciences and by the Hungarian Scientific 
Research Foundation (Grant No.\ OTKA T032981).
\end{acknowledgments} 

%%%%%%%%%%%%%%%%%%%%%%%%%%%%%%%%%%%%%%%%%%%%%%%%%%%%%%%%%%%%%%%%%%%%%%%%%%%
%                             REFERENCES
%%%%%%%%%%%%%%%%%%%%%%%%%%%%%%%%%%%%%%%%%%%%%%%%%%%%%%%%%%%%%%%%%%%%%%%%%%%


\begin{thebibliography}{99}
\bibitem{orig} It is worth noting that the FF was derived originally 
for a different quantity, the phase space contraction rate.
However, it is clear that the physical significance of the FF 
strongly depends on the fact that in various models it can be 
adopted to the entropy production rate as well.
\bibitem{rem-Gauss} We also note that if $\Xi_\tau(x)$ were Gaussian, 
then Eq.~(\ref{flform}) would be satisfied trivially on the condition 
that $\lim_{\tau\to\infty} 2m_\tau/\sigma_{\tau}^{2}=\tau$, 
where $m_\tau$ and $\sigma_{\tau}^{2}$ denote the first and second centered 
moments of $\Xi_\tau(x)$.
\end{thebibliography}
\end{document}